\documentclass{iopart}
\usepackage[T1]{fontenc}
\usepackage[latin1]{inputenc}
\usepackage{bm}
\usepackage{iopams}
\usepackage{setstack}
\usepackage{amssymb}
\usepackage{mathrsfs}
\usepackage{latexsym}
\usepackage{amsfonts}
\usepackage{epsfig}
\usepackage{psfrag}

\newcommand{\nn}{\nonumber\\}

\newcommand{\na}{\mbox{\boldmath$\nabla$}}
\newcommand{\bea}{\begin{eqnarray}}
\newcommand{\ea}{\end{eqnarray}}
\newcommand{\bmult}{\begin{multline}}
\newcommand{\emult}{\end{multline}}
\newcommand{\eea}{\end{eqnarray}}
\newcommand{\ord}{{\cal O}}
\newcommand{\eqref}[1]{(\ref{#1})}

\newcommand{\D}[0]{\underline{\underline{\mathcal D}}}

\usepackage{color}

\begin{document}

\title[Time-resolved density correlations as probe of squeezing]
	{Time-resolved density correlations as probe of squeezing
	in toroidal Bose-Einstein condensates}

\author{Michael Uhlmann}  

\address{
Department of Physics and Astronomy, University of British Columbia, 
Vancouver British Columbia, V6T 1Z1 Canada}

\begin{abstract}
I study the evolution of mean field and linear quantum fluctuations
in a toroidal Bose-Einstein condensate, whose interaction strength
is quenched from a finite (repulsive) value to zero.
The azimuthal {\em equal-time} density-density correlation function
is calculated and shows temporal oscillations with twice the (final)
excitation frequencies after the transition.
These oscillations are a direct consequence of positive and negative
frequency mixing during non-adiabatic evolution.
I will argue that a {\em time-resolved} measurement of the
{\em equal-time} density correlator might be used to calculate the
moduli of the Bogoliubov coefficients and thus the amount of squeezing
imposed on a mode, i.e., the number of atoms excited out of the
condensate.
\end{abstract} 


\maketitle

\section{Introduction}

Bose-Einstein condensates in toroidal traps provide an interesting
opportunity to study superfluidity and other quantum effects in
backgrounds with nontrivial topology
\cite{Bradley,Solenov,Piazza,Piazza2,Jackson,Tsuchiya,Tsuchiya2,Watanabe,Didier,
Lu-Dac,Kanamoto,Kavoulakis,Oegren,Salasnich,Javanainen,Benakli,Petrosyan}.
These geometries can be created ({\em inter alia}) by shining a
blue-detuned laser onto a harmonically trapped condensate thus
creating an effectively repulsive Gaussian core.
The energy minimum is shifted away from the centre of the trap to a
finite radius $r_0$ typically of order $100\,\mu$m
\cite{Andersen,Ryu,Schnelle}.
In present experiments \cite{Andersen,Ryu,Schnelle}, radial trapping
is relatively weak so that azimuthal and radial degrees of freedom
must be considered in general.
Using different trapping techniques, e.g., higher Laguerre-Gaussian
beams as proposed in \cite{Wright}, tighter radial confinement could be
achieved restricting motion in that direction to the trap ground state
and thus making the system effectively one-dimensional.

In the azimuthal direction, the condensate is usually (almost)
homogeneous and obeys periodic boundary conditions permitting
stationary solutions with persistent current, i.e., non-zero
phase gradient, which could be excited through stirring with a laser
\cite{Raman,Madison} or orbital angular momentum transfer from
a Laguerre-Gaussian beam \cite{Andersen,Ryu,Simula}.
By introducing an azimuthal position dependent potential, i.e.,
putting some small obstacles into the torus, inhomogeneous flow
profiles can be generated.
In regions with higher potential, the condensate density will be lower
and the velocity higher possibly leading to a violation of the
Landau criterion and thus an instability if the local speed of sound
is smaller than the flow velocity.
For sufficiently high potentials, the barrier can be overcome only
through tunneling -- a realization of the boson Josephson junction,
where superfluids with different phases are on either side of an
(impenetrable) tunnel barrier
 \cite{Solenov,Piazza,Tsuchiya,Tsuchiya2,Watanabe,Didier,Leszczyszyn}

Local violation of the Landau criterion can also be understood in the
context of cosmic analogs.
As pointed out by Unruh \cite{Unruh}, sound waves in irrotational
fluids obey exactly the same evolution equations as massless
scalar fields in a certain space-time metric, where the effective
curvature is generated by the fluid flow, see \cite{Review} for
review.
The transition from sub- to supersonic flow and {\em vice versa}
can be interpreted as sonic horizons.
It becomes thus, in principle, possible to study some aspects of
cosmic quantum effects in the laboratory, e.g., analogs of Hawking
radiation
\cite{Hawking,BirrellDavies,Unruh,Review,Garay,Jain,Leonhardt,Recati,Carusotto,
Barcelo,Giovanazzi,Wuester,Wuester2,Giovanazzi2,Balbinot,Barcelo3,Macher},
or the freezing and amplification of (quantum) fluctuations in
expanding spacetimes \cite{Guth,BirrellDavies,LiddleLyth,
Review,ich-njp,ich-pra,Barcelo,Fedichev,Fedichev2,Jain2,Barcelo4,Barcelo5,Kurita,Fischer2},
where the latter could be achieved by varying confinement, thus making
the condensate expand
\cite{ich-njp,Fedichev,Fedichev2,Imambekov,Manz}, or by changing the
two-body interaction strength near a Feshbach resonance
\cite{Feshbach}, i.e., reducing the speed of sound
\cite{Barcelo,Jain2,Barcelo4,ich-pra,Kurita,Fedichev2,Fischer2,Carusotto2}.

In this Article, I will concentrate on the latter case and consider
the evolution of a Bose-Einstein condensate in an isotropic toroidal
trap when decreasing the (repulsive) two-body interactions.
Initially, radial confinement shall be relatively weak, i.e., of the
same order as the chemical potential, such that more than one radial
mode is occupied and the system is two-dimensional.
An adiabatic reduction of the coupling strength would lead to a lower
chemical potential, so that the condensate would eventually
become effectively one-dimensional.
For rapid variations of the interactions, however, the condensate --
classical mean field as well as (linear) fluctuations -- might not be
able to follow these changes; the system would not stay in its
ground state throughout the evolution.
The classical order parameter will be homogeneous in the azimuthal
direction due to isotropy of the trap, such that any mean field
excitations can only be in the radial direction, e.g., as radial
breathing motions.
Hence, any azimuthal excitations must originate from the linear
(quantum and thermal) fluctuations such that the symmetry of the
considered trap allows for an unambiguous discrimination between
(classical) mean field and linear quantum fluctuations.

The squeezing of quantum and thermal fluctuations during non-adiabatic
evolution is a very generic phenomenon occurring in many different
physical setting such that toroidal Bose-Einstein condensates
might serve as quantum simulator, e.g., for aspects of cosmic
inflation \cite{Guth,LiddleLyth}.
%
%
During inflation, the size of the universe rapidly increased by a huge
factor thus stretching the wavelength of any excitation mode.
At some point, the quantum fluctuations could not follow the rapid
expansion any more;
they froze and got amplified, an imprint of which is observable as
small anisotropies in the cosmic microwave background -- similar to
the small density ripples in Bose-Einstein condensates.

Although in both cases quantum fluctuations get squeezed during
non-adiabatic evolution, there exist also a few differences between
(massless) quantum fields in expanding space-times and Bose-Einstein
condensates with decreasing interaction strength:
Most notably, the excitation frequencies go to zero in the former case
such that the quantum fluctuations truly freeze, whereas the
excitation energies are always non-vanishing in the latter case and
are generally determined by the confinement of the condensate, e.g.,
the torus radius.
Since, for non-zero quasi-particle frequency, the two-point function
of a squeezed state oscillates with twice the excitation frequency,
such oscillations in the azimuthal density correlations would provide
a signature of the coherent quasi-particle pairs created during the
non-adiabatic evolution.

To exemplify this, let me consider the squeezed harmonic oscillator:
Canonical position, 
$\hat x = 1/\sqrt{2\omega}(\hat a + \hat a^\dagger)$,
and momentum,
$\hat p = - i \sqrt{\omega/2} (\hat a - \hat a^\dagger)$,
can be expressed through raising and lowering operators,
$\hat a^\dagger$ and $\hat a$, whose time-dependences are just given
by oscillating phase factors, e.g.,
$\hat a(t) = e^{-i\omega t}\hat a(0)$.
If the oscillator is in a squeezed state, the vacuum is not defined
with respect to $\hat a$ but rather with respect to a different
operator $\hat b$, i.e., $\hat b|0\rangle = 0$, which can be linked
to the operators $\hat a(0)$ through a Bogoliubov transformation,
$\hat a = \alpha \hat b + \beta^* \hat b^\dagger$.
Obviously, the variances of position and momentum now become
time-dependent, e.g.,
$\langle \hat x^2 \rangle \propto |\alpha|^2 +
|\beta|^2 + 2 |\alpha\beta|\cos(2 \omega t)$.
Thus, a {\em time-resolved} measurement of $\langle\hat x^2\rangle(t)$
would yield the coefficients from which the moduli of the Bogoliubov
coefficients, $|\alpha|$ and $|\beta|$, can be easily inferred and
thus the amount of squeezing, i.e., $|\beta|^2$, can be quantified.
Similarly, the time-dependence of the azimuthal {\em equal-time}
density correlations might serve as probe for
squeezing in toroidal Bose-Einstein condensates.
Note that this
is different from the dynamical structure function, which
is the Fourier transformed of the two-point function at
{\em different} times.

This Article is organized as follows:
In the next section, I review the field equations and their
linearization, before introducing quasi-particles in
Sec.\ \ref{Sec_Quasiparticles}.
Density-density correlations as observable will be discussed there too
with particular emphasis on the time-dependence of the two-point
function due to squeezing of the density fluctuations.
The employed trap geometry is briefly review in Sec.\ \ref{Sec_Torus}
before I consider two explicit examples in Sec.\ \ref{Sec_Crossover}.
For realistic trap parameters, the time-dependent density-density
correlations are calculated for a sudden quench as well as a smooth
$\tanh$-shaped transition of the interaction strength from a finite
value to zero.
The results will be summarized and discussed in
Sec.\ \ref{Sec_Conclusions}.

\section{Field equations}
\label{Sec_Field_equations}

Trapped Bose-Einstein condensate can be described by the interacting
Schr\"odinger field equation (in units where $\hbar = 1$)
\cite{PitaevskiiStringari,Bloch}
    \bea
      i \partial_t \hat \Psi \, = \,
      \left[ - \frac{\na^2}{2m} + V(\bm r)
        + U(t) \hat \Psi^\dagger \hat \Psi
        \right] \hat \Psi \,,
      \label{field_eq}
    \ea
with external potential $V(\bm r)$ and two-body interaction
strength $U(t)$.
Well below transition temperature, most atoms condense in the
lowest state, which acquires a macroscopic occupation number.
It is therefore convenient to split the field operator $\hat\Psi$
into a macroscopic condensate part, which can be treated classically,
and small quantum fluctuations \cite{meanfield}
    \bea
    \hat \Psi \, = \,
    \left(
    \Psi + \hat \chi + \hat \zeta
    \right) \frac{\hat A}{\sqrt{\hat N}} \,,
    \label{meanfield}
    \ea
where $\hat N = \hat A^\dagger \hat A$ counts the total atom number.
The atomic operator $\hat A$ commutes with linear and higher-order
quantum fluctuations, so that particle number is always conserved.
In expansion \eqref{meanfield}, the order parameter
$\Psi = \ord(\sqrt{N})$ describes the condensed atoms,
$\hat \chi = \ord(N^0)$ are linear quantum fluctuations, and
$\hat \zeta \ll \ord(N^0)$ higher orders.

Setting formally $U = \ord(1/N)$, I can expand the field equation
\eqref{field_eq} in powers of $N$ and obtain the Gross-Pitaevskii
equation for the order parameter $\Psi$ \cite{Gross}
    \bea
    i \partial_t \Psi
    \, = \,
    \left( - \frac{\na^2}{2m} + V(\bm r) + U(t) |\Psi|^2
    \right) \Psi \,.
    \label{GP}
    \ea
For the linear fluctuations $\hat \chi$ follows the 
Bogoliubov-de Gennes equation \cite{Bogoliubov}
    \bea
    i \partial_t \hat \chi
    \, = \,
    \left(
    - \frac{\na^2}{2m} + V(\bm r) + 2 U(t) |\Psi|^2
    \right) \hat \chi
    + U(t) \Psi^2 \hat \chi^\dagger \,,
    \label{BdG}
    \ea
which contains a coupling between $\hat \chi$ and its adjoint
$\hat \chi^\dagger$.
The equation of motion for higher-order fluctuations $\hat \zeta$
reads
    \bea
    i \partial_t \hat \zeta
    \, & = & \,
    \left(
    - \frac{\na^2}{2m} + V(\bm r) + 2 U(t)|\Psi|^2
    \right) \hat \zeta
    + U(t) \Psi^2 \hat \zeta^\dagger
    \nn
    && \,
    + U(t) \left(
    2 \Psi \hat \chi^\dagger \hat \chi
    + \Psi^* \hat \chi^2
    + \hat \chi^\dagger \hat \chi^2
    \right)
    + \ord[ U(t) \hat \zeta]
    \label{zeta}
    \ea
with terms quadratic and cubic in $\hat \chi$ in the second line 
acting as source for $\hat \zeta$.
These higher orders $\hat\zeta$ need to be small for the mean-field
expansion \eqref{meanfield} to be valid.
Throughout the rest of this Article, I will not regard Eq.\
\eqref{zeta} any further and concentrate on Eqs.\ \eqref{GP}
for the background and \eqref{BdG} for the linear quantum
fluctuations.

Alternatively to the Gross-Pitaevskii equation \eqref{GP}, the
evolution of the condensate can be described through Bernoulli and
continuity equations for mean field density, $\varrho_0 = |\Psi|^2$,
and phase, $S_0 = \arg\Psi$.
The corresponding linear quantum fluctuations are
    \bea
    \delta\hat\varrho = \Psi^*\hat\chi + \Psi\hat\chi^\dagger
    \,,\qquad
    \delta\hat S = (1/2i\varrho_0)(
    \Psi^*\hat\chi-\Psi\hat\chi^\dagger)
    \label{fluidfluct}
    \ea
and kinematics of the phase fluctuations $\delta\hat S$ would, in the
low energy limit, obey the same evolution equations as a scalar field
in a certain curved space-time, see, e.g.,
\cite{Unruh,Review,Garay,Jain,Leonhardt,Recati,Carusotto,Barcelo,
Giovanazzi,Wuester,Wuester2,Giovanazzi2,Balbinot,Barcelo3,Macher,
ich-pra,ich-njp,Fedichev,Fedichev2,Jain2,Barcelo4,Barcelo5,Kurita,
Fischer2}
-- tools and techniques from general relativity might be applied.
However, in view of the factor $1/\varrho_0$ appearing in the
expression for $\delta\hat S$, linear phase fluctuations are only
well-defined for sufficiently large background densities $\varrho_0$
-- which must evidently fail near the boundary of a trapped condensate
as the smallness of $\delta\hat S$ cannot be guaranteed.
%


Nonetheless, it is still possible to introduce self-adjoint operators
    \bea
    \hat \chi_+ \, = \,
    \frac{1}{\sqrt{2}} \left(
    \hat\chi + \hat\chi^\dagger
    \right)
    \,,\qquad
    \hat \chi_- \, = \,
    \frac{1}{\sqrt{2}i}\left(
    \hat\chi - \hat\chi^\dagger
    \right)
    \label{chipm}
    \ea
similar to those in Eq.\ \eqref{fluidfluct} but without prefactors
$\sqrt{\varrho_0}^{\pm1}$ since these would eventually void
linearization of phase and density in regions with small
(background) density.
The evolution equations for $\hat\chi_\pm$ are most conveniently
written by grouping the linear fluctuation operators into a two-vector
(see also \cite{Leonhardt})
    \bea
    \partial_t\left(\begin{array}{cc}
      \hat \chi_+ \\
      \hat \chi_-
    \end{array}\right)
    \, = \,
    \left( \begin{array}{cc}
      \mathcal C & \mathcal A\\
      - \mathcal B & - \mathcal C
    \end{array}\right)
    \left(\begin{array}{c}
      \hat\chi_+\\
      \hat\chi_-
    \end{array}\right)
    \, = \, \D
    \left(\begin{array}{c}
      \hat\chi_+\\\hat\chi_-
    \end{array}\right) \,,
    \label{eomchipm}
    \ea
where the coefficients of the $2\times2$ matrix $\D$
    \bea
    \mathcal A(\bm r,t) \, & = & \,
    - \frac{\na^2}{2m} + V(\bm r,t) + 2U(t) |\Psi|^2 - U(t) \Re\Psi^2
    \,,\nn
    \mathcal B(\bm r,t) \, & = & \,
    - \frac{\na^2}{2m} + V(\bm r,t) + 2U(t) |\Psi|^2 + U(t) \Re\Psi^2
    \,,\nn
    \mathcal C(\bm r,t) \, & = & \,
    U(t) \Im\Psi^2
    \label{chipmops}
    \ea
are self-adjoint Hilbert-space operators [acting on
$L^2(\mathbb R^D)$] depending on trap potential $V(\bm r,t)$,
interaction strength $U(t)$, as well as classical background
$\Psi(\bm r,t)$.
In \eqref{chipmops}, $\Re\Psi^2$ and $\Im\Psi^2$ are real and
imaginary parts of $\Psi^2$, respectively.

\section{Quasi-particles and observables}
\label{Sec_Quasiparticles}

\subsection{Quasi-particles}

In order to allow for an unambiguous definition of quasi-particles and
thus also the vacuum state, I will assume that the condensate is
initially in a stationary state, i.e., that the order parameter
performs trivial oscillations, $\Psi(t) = e^{-i\mu t}\Psi(0)$ with
chemical potential $\mu$.
Quasi-particles can then, in principle, be defined as eigenmodes of
Eqs.\ \eqref{eomchipm}, i.e., by diagonalizing $\D$, see also
\cite{nonsymmop}, though one needs to be careful because of the
explicit appearance of real and imaginary parts of $\Psi^2$ in the
field equations:
Even if the order parameter performs only trivial oscillations,
$\Psi(t) = e^{-i\mu t}\Psi(0)$, the coefficients of
Eq.\ \eqref{eomchipm} would be time-dependent and an instantaneous
diagonalization of $\D$ would not yield the proper fluctuation
eigenmodes.
It is therefore necessary to absorb the time-dependence of the initial
phase in the definition of the quantum fluctuations (which is
implicitly done in the fluid-dynamic description) before diagonalizing
their evolution equations, see \cite{Leonhardt}.
(Note, however, that the order parameter might retain a
space-dependent phase.)

Following Ref.\ \cite{Leonhardt}, the linear field operators
$\hat\chi_\pm$ can be expanded in terms of initial quasi-particle
solutions (this expansion can also be derived the other way round
starting from the eigenfunctions of $\D$ \cite{nonsymmop})
    \bea
    \hat \chi_+(\bm r,t) & = &
    \mathfrak u_\lambda(\bm r,t) \hat a_\lambda +
    \mathfrak u_\lambda^*(\bm r,t) \hat a_\lambda^\dagger
    \,,\nn
    \hat \chi_-(\bm r,t) & = &
    \mathfrak v_\lambda(\bm r,t) \hat a_\lambda +
    \mathfrak v_\lambda^*(\bm r,t) \hat a_\lambda^\dagger \,,
    \label{quasipart}
    \ea
where each eigenvalue pair $\pm\lambda$ of $\D$ is summed only once.
The entire space-time dependence of $\hat\chi_\pm$ is now
contained in the Bogoliubov functions $\mathfrak u_\lambda(\bm r,t)$
and $\mathfrak v_\lambda(\bm r,t)$, while the mode operators
$\hat a_\lambda$ and $\hat a_\lambda^\dagger$, annihilating or
creating an initial particle with energy $|\lambda|$, are
time-independent.
The initial values of the Bogoliubov functions $\mathfrak u_\lambda$
and $\mathfrak v_\lambda$ are (for stable modes, i.e., purely
imaginary eigenvalues $\lambda$) given by the right eigenfunctions of
evolution equations \eqref{eomchipm}, after absorbing the initial
phase oscillations of the background $\Psi$ into the linear operators
to render the coefficients of \eqref{eomchipm} time-independent.

\subsection{Observables}
\label{Sec_Observables}

The calculation of observables is now straight forward.
Using Eqs.\ \eqref{chipm} and \eqref{quasipart}, density and phase
fluctuations \eqref{fluidfluct} can be expressed in terms of
Bogoliubov functions, $\mathfrak u_\lambda$ and $\mathfrak v_\lambda$,
and mode operators $\hat a_\lambda$, $\hat a_\lambda^\dagger$
    \bea
    \fl\qquad
    \delta\hat\varrho \, = \,
    \sqrt{2} \left(\Re\Psi \hat\chi_+ + \Im\Psi \hat\chi_- \right)
    \, = \,
    \sqrt{2}\left( \mathfrak u_\lambda \Re\Psi
    + \mathfrak v_\lambda \Im\Psi \right)
    \hat a_\lambda + {\rm H.c.}
    \,,\nn\fl\qquad
    \delta\hat S \, = \,
    \frac{1}{\sqrt{2}\varrho}\left(
    \Re\Psi\hat\chi_- - \Im\Psi\hat\chi_+ \right)
    \, = \,
    \frac{1}{\sqrt{2} \varrho}
    \left( \mathfrak v_\lambda \Re\Psi
    - \mathfrak u_\lambda \Im\Psi\right)
    \hat a_\lambda + {\rm H.c.} \,.
    \label{densityphasefluct}
    \ea
Their expectation values $\langle\delta\hat\varrho\rangle$ and
$\langle\delta\hat S\rangle$ must be zero, as they measure only the
deviation from the mean, $\varrho_0$ and $S_0$, respectively.
But their correlations are usually non-zero.
For the density correlations, I obtain
    \bea\fl
    \frac{\left\langle \hat\varrho(\bm r)
    \hat\varrho(\bm r') \right\rangle}
         {\langle\varrho(\bm r)\rangle\langle\varrho(\bm r')\rangle}
    \, = \, 1 + 2
    \frac{\mathfrak u_\lambda(\bm r)\Re\Psi(\bm r)
      + \mathfrak v_\lambda(\bm r) \Im\Psi(\bm r)}
         {|\Psi(\bm r)|^2}
    \frac{\mathfrak u_\lambda^*(\bm r') \Re\Psi(\bm r')
      + \mathfrak v_\lambda^*(\bm r')\Im\Psi(\bm r')}
      {|\Psi(\bm r')|^2} \,,
      \label{densitycorrelations}
    \ea
and for the phase correlations follows
    \bea\fl
    \left\langle \hat S(\bm r) \hat S(\bm r')
    \right\rangle
    \, = \, 1 + \frac{1}{2}
    \frac{\mathfrak v_\lambda(\bm r) \Re\Psi(\bm r)
      - \mathfrak u_\lambda(\bm r)\Im\Psi(\bm r)}
         {|\Psi(\bm r)|^2}
    \frac{ \mathfrak v_\lambda^*(\bm r')\Re\Psi(\bm r')
      - \mathfrak u_\lambda^*(\bm r')\Im\Psi(\bm r')}
         {|\Psi(\bm r')|^2}
    \,,
    \ea
where I took the expectation values with respect to the vacuum state
$|0\rangle$ defined through
$\hat a_\lambda |0\rangle = 0$ $\forall \hat a_\lambda$.
Phase correlations could be measured in interference experiments after
splitting the condensate \cite{Simsarian,Hellweg}.
Since this might be difficult to accomplish due to the nontrivial
topology of the toroidal condensate, I will focus on density
correlations as they can be easily obtained from absorption
images 
\cite{Mathey,Manz,Imambekov,ich-pra,Negretti,Altman,Sykes,Esteve,Schellekens,Modugno,Gangardt}.

\subsection{Time-dependent correlations}

Although in Eq. \eqref{densitycorrelations}, any time arguments were
omitted, for a gapped excitation spectrum, the correlation function
usually performs temporal oscillations even after the external driving
ceased.
As will be shown in the following, this residual time-dependence can
be linked to phase coherence of squeezed excitations and could thus
serve as probe to detect non-adiabatic evolution.
For simplicity, I will assume that the condensate settles into a
stationary state again after some time $t_{\rm end}$ and thus
facilitates the introduction of final quasi-particles.
Analogous to Eq.\ \eqref{quasipart}, the linear field operators
    \bea
    \hat\chi_+(\bm r,t)
    & = &
    \mathfrak f_\kappa(\bm r,t) \hat b_\kappa
    + \mathfrak f_\kappa^*(\bm r,t) \hat b_\kappa^\dagger
    \,,\nn
    \hat\chi_-(\bm r,t)
    & = &
    \mathfrak g_\kappa(\bm r,t) \hat b_\kappa
    + \mathfrak g_\kappa^*(\bm r,t) \hat b_\kappa^\dagger \,,
    \label{quasipartfinal}
    \ea
can be expanded in terms of final quasi-particle solutions
with creation and annihilation operators
$\hat b_\kappa^\dagger$ and $\hat b_\kappa$.
They are related to initial creation and annihilation operators
through a Bogoliubov transformation,
$\hat b_\kappa = \alpha_{\kappa\lambda} \hat a_\lambda +
\beta_{\kappa\lambda}^* \hat a_\lambda^\dagger$, which implicates for
the mode functions
$\mathfrak u_\lambda = \mathfrak f_\kappa \alpha_{\kappa\lambda}
+\mathfrak f_\kappa^* \beta_{\kappa\lambda}$ and
$\mathfrak v_\lambda = \mathfrak g_\kappa \alpha_{\kappa\lambda}
+\mathfrak g_\kappa^* \beta_{\kappa\lambda}$.
The density fluctuations then assume the form
$\delta\hat\varrho =
[h_\kappa(\bm r) e^{-i|\kappa|t} \alpha_{\kappa\lambda} +
h_\kappa^*(\bm r) e^{i|\kappa| t} \beta_{\kappa\lambda}]
\hat a_\lambda + {\rm H.c.}$,
where the function
$h_\kappa(\bm r) = \mathfrak f_\kappa(\bm r) \Re\Psi(\bm r)
+ \mathfrak g_\kappa(\bm r) \Im\Psi(\bm r)$
describes the space-dependence of the final density fluctuation modes
but also contains constant factors accounting, e.g., for the initial
state, or the vacuum amplitude of the density correlations.
Due to the mixing of positive and negative frequency solutions as well
as the mixing of different modes during the non-adiabatic evolution,
the coefficient of the annihilation operator has many contributions
some of which oscillate with positive frequencies $|\kappa|$ and
some with negative frequencies $-|\kappa|$.
Hence, the equal-time correlator (summation over $\kappa$, $\sigma$,
and $\lambda$)
    \bea\fl
    \left\langle \delta\hat\varrho(\bm r)
    \delta\hat\varrho(\bm r') \right\rangle
    & = &
    e^{-i(|\kappa|-|\sigma|)t}
    h_\kappa h_\sigma^*
    \alpha_{\kappa\lambda}\alpha_{\sigma\lambda}^*
    + e^{+i(|\kappa| - |\sigma|)t}
    h_\kappa^* h_\sigma
    \beta_{\kappa\lambda}\beta_{\sigma\lambda}^*
    \nn\fl&&
    + e^{-i (|\kappa| + |\sigma|) t}
    h_\kappa h_\sigma
    \alpha_{\kappa\lambda}\beta_{\sigma\lambda}^*
    + e^{+i(|\kappa| + |\sigma|) t}
    h_\kappa^* h_\sigma^*
    \beta_{\kappa\lambda} \alpha_{\sigma\lambda}^*
    \label{tdcorr}
    \ea
generally comprises many oscillating terms with frequencies
$\pm |\lambda| \pm |\kappa|$.
Bearing in mind that $h_\kappa(\bm r)$ and $h_\sigma(\bm r')$ describe
the space-dependence of the final density fluctuation eigenmodes, any
$\kappa$, $\sigma$ components of \eqref{tdcorr}, e.g.,
$h_\kappa h_\sigma^*\alpha_{\kappa\lambda}\alpha_{\sigma\lambda}^*$
or
$h_\kappa h_\sigma \alpha_{\kappa\lambda}\beta_{\sigma\lambda}^*$,
can be obtained by a suitable (spatial and temporal) projection
of the (measured) correlation function.
After some algebra, the moduli of the Bogoliubov
coefficients $|\alpha_{\kappa\lambda}|$ and $|\beta_{\kappa\lambda}|$
and the constant contributions to $h_\kappa$ follow.

However, since the general calculation is rather tedious, I will
discuss the simpler case with diagonal Bogoliubov coefficients
$\alpha_{\kappa\lambda}$,
$\beta_{\kappa\lambda} \propto \delta_{\kappa\lambda}$ here.
Different eigenmodes do not couple and the density correlator
\eqref{tdcorr} assumes the form
    \bea\fl\qquad
    \langle\delta\hat\varrho(\bm r) \delta\hat\varrho(\bm r')
    \rangle(t)
    \, = \,
    h_\kappa(\bm r) h_\kappa^*(\bm r')
    \left[ |\alpha_\kappa|^2 + |\beta_\kappa|^2 +
      2 \cos( 2|\kappa|t) |\alpha_\kappa\beta_\kappa| \right] \,,
    \label{tdcorr2}
    \ea
where I omitted the phases of the Bogoliubov coefficients
$\alpha_\kappa$, $\beta_\kappa$ and of the spatial function
$h_\kappa$ in the oscillating term (i.e., absorbed them in the
exponentials).
The moduli of the Bogoliubov coefficients $\alpha_\kappa$ and
$\beta_\kappa$ as well as the modulus of $h_\kappa$
(i.e., the adiabatic contribution) can be easily calculated
from mean value,
$h_\kappa h_\kappa^*( |\alpha_\kappa|^2 + |\beta_\kappa|^2)$, and
amplitude, $2 h_\kappa h_\kappa^* |\alpha_\kappa \beta_\kappa|$,
of the temporal oscillations together with the unitarity relation
$|\alpha_\kappa|^2 - |\beta_\kappa|^2 = 1$.

Thus a time-resolved measurement of the {\em equal-time}
correlation function \eqref{densitycorrelations} provides a direct
means of detecting non-adiabatic evolution of the linear quantum
fluctuations, i.e., quasi-particles squeezed out of the vacuum
can be observed.
Any information about the initial state, e.g., whether it is
thermal or the vacuum, is only encoded in the prefactor
$h_\kappa(\bm r)$ such that a clear discrimination between
amplification and initially present particles is possible
-- a feature the phonon detection scheme proposed in
\cite{Ralf} lacks.
Note that the correlation function \eqref{tdcorr} is at {\em equal}
times.
Its Fourier transformed should therefore not be confused with the
dynamical structure factor
$S(k,\omega) = \int d^Dkd\omega/(2\pi)^{D/2+1} e^{i(\omega t - \bm k\bm r)}
\langle \hat\varrho(\bm r,t) \hat\varrho(\bm 0, 0)\rangle$, cf.\ e.g.\
\cite{PitaevskiiStringari,Cherny}, which is the Fourier transformed of
the density correlations at {\em different} times.

\section{Toroidal condensate}
\label{Sec_Torus}

Inspired by recent experiments \cite{Andersen,Ryu,Schnelle}, I will
consider a harmonic trap of frequency $\omega$ with repulsive
Gaussian core of strength $V_0$ and width $\sigma$
    \bea
    V(r) \, = \,
    \frac{m \omega^2 r^2}{2}
    + V_0 \exp\left\{ -\frac{r^2}{\sigma^2}\right\} \,.
    \label{torus}
    \ea
For simplicity, I will assume that the condensate is
quasi-two-dimensional with radial and azimuthal degrees of freedom.
Due to the isotropy of the trap, any mean field motion can only be
excited in the radial direction, whereas quantum fluctuations occur at
all azimuthal wavenumbers.
After integrating out the radial dependence of the density,
$\hat\varrho(\phi,t) = \int dr r\, \hat\varrho(\phi,r,t)$,
the azimuthal density correlations assume the form
    \bea\fl
    \frac{\left\langle\hat\varrho(\phi,t)
      \hat\varrho(\phi',t)\right\rangle}
         {\left\langle\hat\varrho(\phi,t)\right\rangle
           \left\langle\hat\varrho(\phi',t)\right\rangle}
         \, = \,
         1 + \frac{e^{im(\phi-\phi')}}{\sqrt{2\pi}}
         \frac{|\delta\varrho_m|^2}{\varrho_0^2}
         \, = \,
         1 + \frac{e^{im(\phi-\phi')}}{2\pi} \sum_\lambda
         \frac{|\delta\varrho_m^\lambda|^2}{\varrho_0^2} \,,
    \ea
where
    \bea
    \delta\varrho_m^\lambda \, = \,
    \int\limits_0^{2\pi} d\phi
    \int \limits_0^\infty dr r\, \frac{e^{-im\phi}}{\sqrt{\pi}}
    \left[\mathfrak u_\lambda(\bm r,t) \Re\Psi(\bm r,t)
    + \mathfrak v_\lambda(\bm r,t) \Im\Psi(\bm r,t) \right]
    \ea
are the contributions of the initial quasi-particle modes $\lambda$,
see \ref{App_Basis_expansion} for more details.

\section{Quench}
\label{Sec_Crossover}

In the examples, a condensate of $10^5$ interacting $^{85}$Rb atoms in
a quasi-two-dimensional trap with frequency
$\omega = 40\pi\,{\rm  Hz}$, cf.\ Eq.\ \eqref{torus}, will be
considered.
The Gaussian beam shall have a size of $\sigma= 175\,\mu{\rm m}$
and intensity $V_0 = 5000 \hbar\omega$ so that the potential
minimum of the torus lies at $r_{\rm min} = 277\,\mu$m.
This trap geometry is loosely inspired by those used in Refs.\
\cite{Andersen,Simula}, though it should be noted that I use a
much higher intensity $V_0$ of the Gaussian beam and also a lower
particle number to facilitate a quadratic approximation of the
radial trapping with effective frequency $\omega_{\rm eff}$, see
\ref{App_Basis_expansion} for details.
Because confinement in the third direction ($z$) is supposed to be
sufficiently tight and enters the calculation only via the
effective two-dimensional coupling strength
$U_{\rm 2D} = (2 \sqrt{2\pi}/m) a_s(t)/a_\perp$,
I will not give
any particular value for $\omega_\perp$ nor for the initial
$s$-wave scattering length $a_s$.
However, the general features, i.e., the oscillations of the azimuthal
density correlations, should be retained for three-dimensional trap
geometries, higher particle numbers, or less intense Gaussian cores
as well.

The interaction strength is quenched from 
$U_0 = 0.025 \hbar \omega_{\rm eff} a_{\rm eff}$
to $U_1 = 0$.
The initial chemical potential is
$\mu_{\rm in} = 7.29\,\omega_{\rm eff}$, so that the lowest $9$
radial trap modes will have significant occupation numbers and
the condensate is thus in a quasi-two-dimensional regime.
When adiabatically lowering the nonlinear coupling $U$,
the chemical potential would decrease and the population of
all radial trap modes except the lowest would decline -- the
system would become effectively one-dimensional.
For rapid changes of $U$, this, however, is generally not the
case any more and (radial) breathing oscillations of the mean
field will be excited.
Several radial trap modes will have macroscopic occupation.
Evidentially, as the final state is not stationary, the definition of a
chemical potential is not meaningful any more.

The results presented in the following have been obtained using a
basis expansion of the order parameter, $\Psi$, and the linear
fluctuations, $\hat\chi_\pm$, see \ref{App_Basis_expansion}, and
keeping the lowest 15 radial basis functions.
The initial state was propagated using a fourth-order Runge-Kutta
algorithm with adaptive step width for the coupled evolution equations
of mean field \eqref{GPbasis} and fluctuations \eqref{eomchipmbasis}.


\subsection{Sudden quench}

\begin{figure}[!hbt]
  \psfrag{t}{$t$} 
  \psfrag{n}{$n$}
  \psfrag{dr2}{$\qquad|\delta\varrho_n|^2/\varrho_0^2$}
  \psfrag{0}{$0$}
  \psfrag{2}{$2$}
  \psfrag{4}{$4$}
  \psfrag{6}{$6$}
  \psfrag{8}{$8$}
  \psfrag{10}{$10$}
  \psfrag{20}{$20$}
  \psfrag{0.005}{$0.005$}
  \psfrag{0.01}{$0.01$}
  \psfrag{0.015}{$0.015$}
  \psfrag{0.02}{$0.02$}
  \psfrag{0.025}{$0.025$}
  \includegraphics[width=\textwidth]{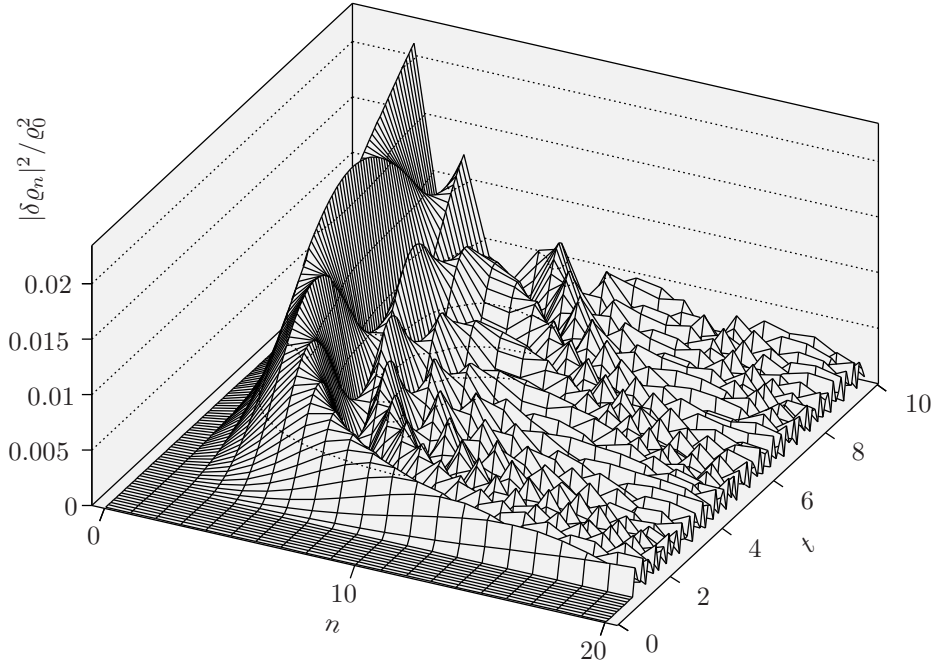}
  \caption{Fourier components $|\delta\varrho_n|^2/\varrho_0^2$ of the
    density correlator as a function of azimuthal mode $n$ and time
    $t$ for a sudden quench of the interaction strength
    from $U_0 = 0.025\,\hbar\omega_{\rm eff}a_{\rm eff}$ to
    $U_1 = 0$ after $t_{\rm sweep} = 1s$.
    \label{sudden_spectrum_evo}
  }
\end{figure}

\begin{figure}[!hbt]
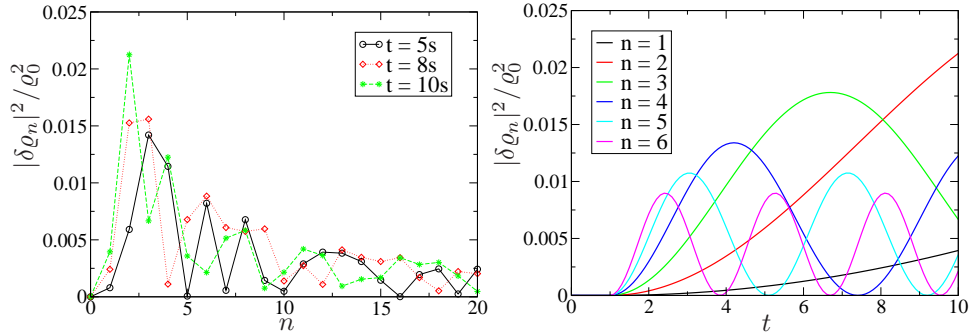

  \begin{minipage}{0.48\textwidth}
    \psfrag{dr2}{$|\delta\varrho_n|^2/\varrho_0^2$}
    \psfrag{n}{$n$}
    \includegraphics[width=\textwidth]{sudden_spectrum.eps}
  \end{minipage}
  \begin{minipage}{0.48\textwidth}
    \psfrag{dr2}{$|\delta\varrho_n|^2/\varrho_0^2$}
    \psfrag{t}{$t$}
    \includegraphics[width=\textwidth]{sudden_drho_evo.eps}
  \end{minipage}
  \caption{Density-density correlations for sudden quench
    from $U_0 = 0.025\,\hbar\omega_{\rm eff} a_{\rm eff}$ to
    $U_1 = 0$ at $t_{\rm sweep} = 1s$.
    On the left, the spectrum is plotted at different times,
    whereas the time dependence of the lowest six components
    $|\delta\varrho_1|^2/\varrho_0^2$ to
    $|\delta\varrho_6|^2/\varrho_0^2$ is shown on the right.
    \label{sudden}
  }
\end{figure}


A sudden reduction of the interaction strength $U$ from a finite
value to zero is probably the easiest quench dynamics that can be
realized in experiments.
In Figure \ref{sudden_spectrum_evo}, the time-dependence of the
azimuthal density correlation spectrum is shown for such a transition
from $U_0 = 0.025 \,\hbar\omega_{\rm eff} a_{\rm eff}$ to $U_1 = 0$.
Relative density correlations on the percent level can be observed
for low azimuthal excitations.
However, the $n = 0$ mode is not squeezed as this would correspond
to fluctuations in the (conserved) total particle number.
The left plot of Figure \ref{sudden} shows the spectrum at different
times, while the right plot displays the time-dependence of the
the lowest azimuthal modes.
The correlation spectrum does not converge to a constant value, but
rather exhibits temporal oscillations well after the quench even
though the interaction strength has been tuned to zero.

These temporal oscillations of correlation functions might seem a bit
puzzling at first but are actually a direct consequence of the
non-adiabatic evolution during the interaction quench together with a
gapped final excitation spectrum, see also Section
\ref{Sec_Observables} especially Eqs.\ \eqref{tdcorr} and
\eqref{tdcorr2}.
The oscillation frequencies observed in the plots are in good
agreement with the expected values:
After the interactions have been tuned to zero, the azimuthal
excitation frequencies are
$\omega_{n} = n^2/2mr_{\rm min}^2 = n^2 \times 0.306\,{\rm Hz}$ above
the corresponding mean field in each radial mode.
For $n = 4$, $n = 5$, and $n = 6$, this implies period times
$T_4 = 6.42\,{\rm s}$, $T_5 = 4.11\,{\rm s}$, and
$T_6 = 2.85\,{\rm s}$ -- the values observed in Figures
\ref{sudden_spectrum_evo} and \ref{sudden}.
For lower excitations, one oscillation period lasts much longer such
that no full cycles of $|\delta\varrho_n|^2(t)$ can be observed during
the propagation time.
Whereas higher excitations undulate very quickly, e.g., $n = 20$ has
repetition time $T_{20} = 0.26\,{\rm s}$, making it much harder to
resolve in Figure \ref{sudden_spectrum_evo} due to finite grid size of
the plot.
Also, the amplitude of the oscillations declines with $n$, as it is
related to the Bogoliubov $|\beta|$ coefficients, which, for a
sudden transition, read
    \bea
    |\beta| \, = \,
    \frac{1}{2}
    \left| \sqrt{\frac{\omega_{\rm out}}{\omega_{\rm in}}}
      -\sqrt{\frac{\omega_{\rm in}}{\omega_{\rm out}}} \right| \,.
      \label{bogosudden}
    \ea
The azimuthal Bogoliubov coefficients, i.e., summed over all radial
excitations, can be calculated from minimal and maximal values of the
oscillations in Fig.\ \ref{sudden} (though it might be difficult to
extract these values from the plots).
From the numerical data, I have for the mode $n = 4$ 
as minimum
$|\delta\varrho_4|^2_{\rm min}/\varrho_0^2 = 2.44\times10^{-6}$
and as maximum
$|\delta\varrho_4|^2_{\rm max}/\varrho_0^2 = 1.34\times10^{-2}$,
such that the Bogoliubov coefficient of that azimuthal
mode is $|\beta_{n = 4}|^2 \approx 18$.
This relatively large number can be understood from the huge energy
difference before and after the quench:
Initially, the condensate is in an interacting regime and the
quasi-particles are phonon-like excitations whose frequency is
dominated by the radial contribution of order of the effective
trap frequency, $\omega_{\rm eff} = 281\,{\rm Hz}$,
cf.\ \ref{App_Basis_expansion}.
After the quench, however, only the small azimuthal $\ord(1\,\rm{Hz})$
part remains as the condensate becomes non-interacting.

The oscillations in the azimuthal equal-time density correlations can
be interpreted as squeezing of quasi-particle (atom) pairs out of the
quantum vacuum.
Due to angular momentum conservation, one of the partners must have
azimuthal wavenumber $+n$ and one $-n$, i.e., one moves clockwise and
one counter-clockwise.
If both quasi-particles are at the same place, they will not contribute
to the non-local correlations, hence the minima of \eqref{tdcorr2}.
As they are moving away from each other, their contribution to
$\langle\hat\varrho(\phi)\hat\varrho(\phi')\rangle$ grows until
it reaches its maximum when both quasi-particles are farthest apart.
A time-resolved measurement of the density-density correlations can
be used to trace these coherent pairs and thus infer the amount of
squeezing incurred during the transition, e.g., the Bogoliubov
coefficients, $|\beta_n|^2$, giving the number of created
quasi-particles in each azimuthal mode (summing over the different
radial components) \cite{BirrellDavies}.


\subsection{$\tanh(\gamma t)$ sweep}

\begin{figure}[!hbt]
  \psfrag{t}{$t$} 
  \psfrag{n}{$n$}
  \psfrag{dr2}{$\qquad|\delta\varrho_n|^2/\varrho_0^2$}
  \psfrag{0}{$0$}
  \psfrag{2}{$2$}
  \psfrag{4}{$4$}
  \psfrag{6}{$6$}
  \psfrag{8}{$8$}
  \psfrag{10}{$10$}
  \psfrag{20}{$20$}
  \psfrag{0.002}{$0.002$}
  \psfrag{0.004}{$0.004$}
  \psfrag{0.006}{$0.006$}
  \psfrag{0.008}{$0.008$}
  \psfrag{0.01}{$0.01$}
  \psfrag{0.012}{$0.012$}
  \includegraphics[width=\textwidth]{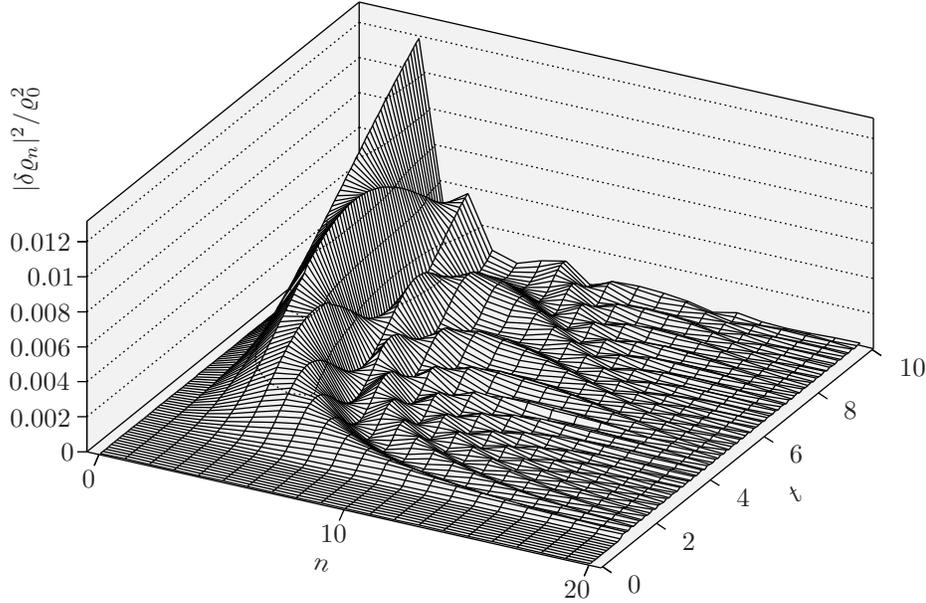}
  \caption{Time-dependence of the correlation spectrum for a smooth
    $\tanh$-shaped transition between
    $U_0 = 0.025\,\hbar\omega_{\rm eff} a_{\rm eff}$ and $U_1 = 0$, cf.\
    Eq.\ \eqref{U_tanh}.
    The transition occurs at $t_{\rm sweep} = 1\,{\rm s}$ with
    sweep rate is equal to the harmonic trap frequency
    $\gamma = \omega = 40\pi\,{\rm Hz}$.
  }
  \label{tanh_spectrum_evo}
\end{figure}

\begin{figure}[!hbt]
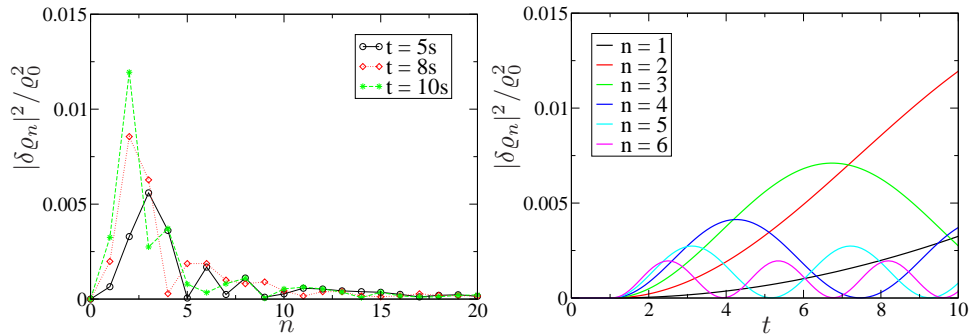

  \begin{minipage}{0.48\textwidth}
    \psfrag{dr2}{$|\delta\varrho_n|^2/\varrho_0^2$}
    \psfrag{n}{$n$}
    \includegraphics[width=\textwidth]{tanh_spectrum.eps}
  \end{minipage}
  \begin{minipage}{0.48\textwidth}
    \psfrag{dr2}{$|\delta\varrho_n|^2/\varrho_0^2$}
    \psfrag{t}{$t$}
    \includegraphics[width=\textwidth]{tanh_drho_evo.eps}
  \end{minipage}
  \caption{Density-density correlations for a smooth $\tanh$
    transition from $U_0 = 0.025\,\hbar\omega_{\rm eff} a_{\rm eff}$
    to $U_1 = 0$ with sweep rate
    $\gamma = \omega = 40 \pi\,{\rm Hz}$ at
    $t_{\rm sweep} = 1\,{\rm s}$, cf.\ Eq.\ \eqref{U_tanh}.
    The left plot shows the spectrum at different times and
    the right shows the time-dependence of the lowest
    azimuthal correlation.
  }
  \label{tanh}
\end{figure}


As second example for the dynamics, I will consider a smooth
transition ($\gamma > 0$)
    \bea
    U(t) \, = \,
    \frac{U_0 + U_1}{2}
    - \frac{U_0 - U_1}{2}
    \tanh\left[\gamma ( t - t_{\rm sweep} )\right]
    \label{U_tanh}
    \ea
from $U_0 = 0.025\,\hbar\omega_{\rm eff} a_{\rm eff}$ to
$U_1 = 0$, i.e., between the same limiting values as before.
The sweep rate $\gamma$ shall be equal to the trap frequency $\omega$,
such that only the low-lying excitations with energies of about the
same order as the trap frequency will evolve non-adiabatically.
The results are plotted in Figures \ref{tanh_spectrum_evo} and
\ref{tanh}.
Again, correlations on the percent level can be observed, though,
as expected, smaller than in the previous example.
Temporal oscillations of $|\delta\varrho_n^2|$, characteristic for
non-adiabatic evolution, can be observed for low excitations only
and subside much more rapidly with increasing $n$ than for a sudden
transition.
The spectrum $|\delta\varrho_n|^2$ becomes (almost) constant in time
for higher-energetic modes $n \approx 15...20$ already because the
excitation energies are smaller than the inverse sweep rate $1/\gamma$
and these higher fluctuations can (at least partially) adapt to
changes of the background.
%



\section{Discussion}
\label{Sec_Conclusions}

In summary, I studied the evolution of quasi-two-dimensional Bose-Einstein
condensates in isotropic toroidal traps when tuning the (repulsive)
interaction strength from a finite value to zero.
Since present experiments \cite{Andersen,Ryu,Schnelle} provide only relatively
weak radial confinement, the initial interaction strength was chosen such that
the condensate is in an effectively two-dimensional regime with radial and
azimuthal degrees of freedom.
In the third direction ($z$), confinement shall be sufficiently tight
such that motion is restricted to the trap ground state and can be
integrated out.

Due to the isotropy of the trap, the classical mean field must be
homogeneous in the azimuthal direction.
Hence, any background motion must occur radially.
The linear quantum fluctuations, on the other hand, are only subject
to (angular) momentum conservation and might thus have non-trivial
azimuthal dependence, which can be treated through a Fourier
transformation.
For each Fourier mode, two coupled one-dimensional field equations
(in the radial direction) with gap $n^2/2mr_{\rm min}^2$ follow, which
can be easily solved numerically.
By integrating out the radial dependence and considering only
the (relative) azimuthal correlations, a clear signature of the quantum
fluctuations can be provided.

The radial dependence of mean field and linear quantum fluctuations
was expanded in terms of harmonic oscillator functions.
For the initial finite interaction strength, the condensate (mean field)
is radially distributed over several modes.
When slowly (adiabatically) reducing the interaction strength $U$,
more and more atoms would gather in the trap ground state,
while the occupation of all higher modes would diminish until the
condensate becomes effectively one dimensional.
For rapid variations of $U$, however, the condensate might not be able to
follow these changes of the interaction strength and will therefore
not stay in its ground state.
A fraction of the atoms would remain in higher trap states and the
mean field would split up into several parts, whose phases oscillate
with different frequencies -- the condensate would undergo classical
oscillations in the radial direction.
It should be noted that, even though the condensate performs breathing
motion and is thus split into different parts evolving with vastly
different frequencies, the mean field is still coherent, i.e., the
phases of the classical excitations and the non-excited bulk
are still uniquely related. 
Coupling to an environment or the back-reaction of quantum
fluctuations might, however, lead to loss of phase coherence so that
the condensate becomes fragmented into several
mutually incoherent parts \cite{Bader,Fischer}.

For non-vanishing interaction strength after the quench, $U \neq 0$,
and if the mean field undergoes radial breathing, the
background-dependent terms in the linear field equations,
$U|\Psi(r,t)|^2$ and $U\Psi^2(r,t)$, would become periodic in time and
could amplify quantum fluctuations through resonance.
Therefore, I considered only the case, where $U$ is tuned to zero and
the theory becomes non-interacting, i.e., the linear excitations
evolve independently of the condensate background.
%
%
In view of the azimuthal homogeneity of the mean field, azimuthal
density correlations -- the radial dependence has
been integrated out -- represent a suitable observable for the quantum
fluctuations independent of any background motion.
Two different time-dependences of the interaction strength were
considered:
firstly, a sudden change from a finite $U_0$ to zero and, secondly,
a smooth $\tanh$-shaped transition between the same values of the
interaction strength.
In both cases, squeezing of quasi-particles could be observed,
which manifests as temporal oscillations in the azimuthal correlation
functions.
This can be understood in a simple picture: due to angular momentum
conservation, quasi-particles are always excited as pairs with
azimuthal wavenumbers $\pm n$.
With one quasi-particle (atom) circling the torus clockwise and the
other counter-clockwise, the oscillations in the correlation function
can then be understood through co-incidence of the pair.
For a sudden transition, all modes evolve non-adiabatically, whereas
a smooth $\tanh$ shape of the interaction strength violates
adiabaticity only for low excitations, while higher modes stay closer
to their ground state.
In both cases, relative (vacuum) density-density correlations on the
percent level could be observed.

Temporal oscillations in the equal-time correlation function after the
quench are closely related to finite quasi-particle frequencies
and non-adiabatic evolution of the fluctuations
-- initial particle solutions oscillating with positive frequency
before the quench usually comprise positive and negative frequency parts
afterwards, where the amount of mixing can be quantified through
Bogoliubov coefficients $\alpha$ and $\beta$.
For non-zero $\beta$, i.e., non-adiabatic evolution, the density
correlator will assume oscillations with twice the final
quasi-particle frequency.
From mean value and amplitude of the oscillations, the Bogoliubov
coefficients as well as the adiabatic correlations can be obtained.
Thus, a {\em time-resolved} measurement of the {\em equal-time}
density-density correlation function permits the determination of the
Bogoliubov coefficients $|\beta|^2$ as signature of quasi-particle
squeezing.

In contrast to methods aiming at the detection of single phonons,
e.g., \cite{Ralf}, the temporal oscillations of the density
correlation are as signature of squeezing relatively robust against
changes of the initial quantum state.
For instance, a thermal occupation number would merely appear as
prefactor to the correlations of a mode, whereas the Bogoliubov
coefficients $|\beta|$ as measure for the amplification of the initial
fluctuations (and quasi-particles) follow from the relative amplitude
compared to the mean of the oscillations, which is unaffected by such
a prefactor.
If the excitation frequencies would go to zero, which happens in
expanding Bose-Einstein condensates after the trap has been turned off
\cite{ich-njp,Imambekov,Manz,Modugno,Gangardt} or during cosmic
inflation \cite{Guth,LiddleLyth}, the density correlations would
converge and a time-resolved measurement would yield no additional
information.
In particular, it would not be possible to distinguish between
contributions stemming from initial particles and those due
to the amplification of fluctuations, i.e.,
squeezing.

A time-dependent measurement of the {\em equal-time} density
correlations is also fundamentally different from
obtaining the dynamical structure factor
\cite{PitaevskiiStringari,Cherny}, which is the temporal Fourier
transformed of the density correlator at {\em different} times.
To exemplify these differences, let me consider a simple one-mode
model $\hat\varrho = f(t) \hat a + f^*(t) \hat a^\dagger$.
Assuming that the system is in squeezed state,
$f(t) = \alpha e^{-i\omega t} + \beta^* e^{i\omega t}$, 
the propagator reads
$\langle \hat\varrho(t) \hat \varrho(t') \rangle = f(t) f^*(t')
= (|\alpha|^2 + \alpha\beta e^{-2i\omega t'}) e^{-i\omega(t - t')}
+ (|\beta|^2 + \alpha^*\beta^*e^{2i\omega t'}) e^{i\omega(t - t')}$
and the dynamical structure factor has two peaks at $\pm \omega$
with time-dependent coefficients.
It is, in principle, possible to obtain the Bogoliubov coefficients
and thus the amount of squeezing from the time-dependence of these
peaks as well, though a time-dependent measurement where {\em both}
times $t$ and $t'$ are varied would be necessary.


\section*{Acknowledgements}

I would like to thank William G.\ Unruh and Uwe R.\ Fischer for
helpful comments and discussions.
This work was supported by the Alexander von Humboldt Foundation and
NSERC of Canada.


\appendix

\section{Basis expansion}
\label{App_Basis_expansion}

\subsection{General basis}

Method of choice for the numerical propagation of mean field $\Psi$
as well as quantum fluctuations $\hat\chi_\pm$ is a basis expansion.
To remain general, I will formally consider non-orthogonal bases in
the following, i.e., tensor notation will be adopted and I will
distinguish between covariant $\{f_\alpha(\bm r)\}$ and contravariant
$\{f^\alpha(\bm r)\}$ bases.
They are normalized
$\langle f^\alpha, f_\beta\rangle_{L^2} = \delta{^\alpha}_\beta$ with
$L^2$ inner product
$\langle f, g \rangle_{L^2} = \int d^Dr f^*(\bm r) g(\bm r)$.
Non-orthogonal basis might occur, e.g., when using the eigenfunctions
of a non-symmetric operator as basis, cf.\ also $\D$ \cite{nonsymmop},
or when picking a particularly suitable numerical basis.

The components of the order parameter and the linear fluctuations
    \bea
    \Psi(\bm t,r)
    \, = \,
    f_\alpha(\bm r)\Psi^\alpha(t)
    \,,\quad
    \hat\chi_\pm(\bm r,t)
    \, = \,
    f_\alpha(\bm r) \hat \chi_\pm^\alpha(t)
    \label{basisexp}
    \ea
follow from their projection onto the covariant basis
$\Psi^\alpha = \langle f^\alpha, \Psi\rangle_{L^2}$ and
$\hat\chi_\pm^\alpha = \langle f^\alpha, \hat\chi_\pm \rangle_{L^2}$.
The Gross-Pitaevskii equation \eqref{GP} reads in basis expansion
    \bea
    i\partial_t \Psi^\alpha
    \, = \,
    \mathcal K{^\alpha}_\beta \Psi^\beta
    + U \mathcal M{^\alpha}_{\beta\gamma\delta}
    (\Psi^\beta)^* \Psi^\gamma \Psi^\delta \,,
    \label{GPbasis}
    \ea
where the entire space-dependence (and possibly the time-dependence
of the external confinement $V$) is integrated out and thus contained
in the coupling tensors
$\mathcal K{^\alpha}_\beta = \int d^Dr (f^\alpha)^* (-\na^2/2m + V ) f_\beta$
and
$\mathcal M{^\alpha}_{\beta\gamma\delta} =
\int d^Dr\, (f^\alpha)^* f_\beta^* f_\gamma f_\delta$.
For the linear fluctuations, I obtain
    \bea
    i \partial_t
    \left(\begin{array}{c}
      \hat \chi_+^\alpha\\
      \hat \chi_-^\alpha
    \end{array}\right)
    \, = \,
    \left(\begin{array}{cc}
      \mathcal C{^\alpha}_\beta & \mathcal A{^\alpha}_\beta \\
      -\mathcal B{^\alpha}_\beta & - \mathcal C{^\alpha}_\beta
    \end{array}\right)
    \cdot
    \left(\begin{array}{c}
      \hat \chi^\beta_+\\
      \hat \chi^\beta_-
    \end{array}\right) \,.
    \label{eomchipmbasis}
    \ea
The coupling tensors
$\mathcal A{^\alpha}_\beta = \int d^Dr\, (f^\alpha)^* \mathcal A
f_\beta$,
$\mathcal B{^\alpha}_\beta = \int d^Dr\, (f^\alpha)^* \mathcal B
f_\beta$, and
$\mathcal C{^\alpha}_\beta = \int d^Dr\, (f^\alpha)^* \mathcal C
f_\beta$
follow by taking the matrix elements of operators \eqref{chipmops}
with respect to the basis $\{f_\alpha\}$ and its dual $\{f^\alpha\}$.
Note that, due to the non-linearity of the original field equation
\eqref{field_eq} and thus the appearance of the mean field
$\Psi$ in the linear evolution equations \eqref{eomchipm} for
$\hat \chi_\pm$, the matrix coefficients in \eqref{eomchipmbasis}
usually contain integrals of four basis functions similar to
$\mathcal M{^\alpha}_{\beta\gamma\delta}$ and summation over the
components of the order parameter $\Psi$.

From Eqs.\ \eqref{quasipart} and \eqref{basisexp}, I can infer
the basis expansion of the Bogoliubov functions $\mathfrak u_\lambda$
and $\mathfrak v_\lambda$
    \bea
    \mathfrak u_\lambda(\bm r,t)
    \, = \,
    f_\alpha(\bm r) \mathfrak u_\lambda^\alpha(t)
    \,,\quad
    \mathfrak v_\lambda(\bm r,t)
    \, = \,
    f_\alpha( \bm r) \mathfrak v_\lambda^\alpha(t)
    \,.
    \label{bogobasis}
    \ea
so that the linear field operators $\hat\chi_\pm$ can be
described by time-independent operators $\hat a_\lambda^\dagger$,
$\hat a_\lambda$ creating or annihilating an initial quasi-particle
with energy $|\lambda|$ and time-dependent coefficients
$\mathfrak u_\lambda^\alpha(t)$ and $\mathfrak v_\lambda^\alpha(t)$.
The initial values of the Bogoliubov functions are given by the (up
and down) components of the right eigenvectors of the coupling matrix
appearing in \eqref{eomchipmbasis}, and can be propagated using
Eq.\ \eqref{eomchipmbasis}.
As the evolution equation is first order in time, positive and
negative frequency solutions appear separately in the spectrum and one
needs to be careful not to double count the pair, i.e., (for stable
modes) eigenvalues with negative imaginary part correspond to
annihilation operators $\hat a_\lambda$ and those with positive
imaginary part yield creators $\hat a_\lambda^\dagger$.
Inserting expansions \eqref{basisexp} and \eqref{bogobasis} into
\eqref{densitycorrelations}, observables such as the density
correlations can be calculated.
One should, however, bear in mind that the basis functions $f_\alpha$
are usually not real and thus complex conjugation of the coefficients
$\mathfrak u_\lambda^\alpha$ etc.\ does not yield the coefficients of
$(\mathfrak u_\lambda^\alpha)^*$ etc.\ with respect to the basis
$\{f_\alpha\}$ but rather those with respect to $\{f_\alpha^*\}$.


\subsection{Toroidal condensates}

Toroidal Bose-Einstein condensates can be created using potential
\eqref{torus}, which has its minimum at
$r_{\rm min}^2 = \sigma^2 \ln( 2V_0/m\omega^2\sigma^2) > 0$
for $2 V_0 > m\omega^2 \sigma^2$.
In order to simplify the analysis a bit, this potential can be
expanded to second order about this minimum
    \bea
    V(r) \, = \,
    V(r_{\rm min}) +
    \frac{m \omega_{\rm eff}^2}{2} (r - r_{\rm min})^2
    + \ord[ (r - r_{\rm min})^3] \,,
    \label{radialharm}
    \ea
and an effectively harmonic potential with frequency
$\omega_{\rm eff}^2 = 2 \omega^2 \ln(2V_0/ m \omega^2\sigma^2)
= 2 \omega^2 r_{\rm min}^2/\sigma^2$ can be obtained.
The first term, $V(r_{\rm min})$, yields only a constant energy shift
and will be omitted in the following.
This expansion \eqref{radialharm} of the potential can be used
if the radial extend $\Delta r$ of the condensate is much smaller than
the torus radius $r_{\rm min}$,
which can be achieved by demanding that the harmonic oscillator 
length $a_{\rm eff} = 1/\sqrt{m\omega_{\rm eff}}$ be much smaller than
$r_{\rm min}$ and only the lowest few radial modes being occupied,
i.e., a chemical potential of order $\omega_{\rm eff}$.


In view of the symmetries of the trap -- the potential depends only
on $r = |\bm r|$ and is independent of the azimuthal angle $\phi$ --
it is advantageous to change to planar polar coordinates,
$\bm r = r \bm e_r(\phi)$, and to introduce bases for radial and
azimuthal dependence.
Harmonic oscillator functions $h_\alpha(r - r_{\rm min})$ centered at
$r_{\rm min}$ seem to be an obvious choice for the radial basis.
But one should, however, bear in mind that polar coordinates yield an
additional factor $r$ as measure in the inner product
$\int d^2r h_\alpha^* h_\beta = \int r dr d\phi h_\alpha^* h_\beta$.
It is therefore more convenient to absorb this factor and to use
slightly different functions
    \bea
    f_\alpha( r - r_{\rm min})
    \, = \, \frac{h_\alpha(r - r_{\rm min})}{\sqrt{r}} \,,
    \label{radialbasis}
    \ea
%
which are approximately orthonormal
$\int_0^\infty dr r\,
f_\alpha(r - r_{\rm min})
f_\beta(r - r_{\rm min})
= \delta_{\alpha\beta}$.
Corrections to this orthonormality are exponentially suppressed
$\propto e^{-(\alpha+\beta) a_0/r_{\rm min}}$.
[Of course, this suppression does not hold for higher modes
$\alpha \gtrsim r_{\rm min}/a_0$, but the occupation numbers in
these modes are very small because the condensate is radially
localized near $r_{\rm min}$.
Also, expansion \eqref{radialharm} of the radial potential
\eqref{torus} to second order would break down first.]
%


In view of the isotropy of the potential \eqref{torus}, the azimuthal
dependence of mean field $\Psi$ and linear fluctuations $\hat\chi_\pm$
is most conveniently described in terms of plane waves
    \bea
    g_m(\phi) \, = \,
    \frac{1}{\sqrt{2\pi}} e^{i m \phi}
    \,,\qquad
    m \in \mathbb Z
    \label{azimuthalbasis}
    \ea
obeying periodic boundary conditions, $g_m(\phi) = g_m(\phi + 2 \pi)$,
and normalization
$\int_0^{2\pi} d\phi g_m^*(\phi) g_n(\phi) = \delta_{nm}$.
Due to isotropy of the trap \eqref{torus}, the classical mean field,
$\Psi(r, \phi, t) = \frac{1}{\sqrt{2\pi}} f_\alpha(r) \Psi^\alpha(t)$,
must be independent of the angle $\phi$ and the azimuthal label
on $\Psi^{m=0,\alpha}$ can be omitted.
The Bogoliubov functions
$\mathfrak u_\lambda(\bm r,t) = g_m(\phi) f_\alpha(r) \mathfrak
u_\lambda^{m\alpha}(t)$
and 
$\mathfrak v_\lambda(\bm r,t) = 
g_m(\phi) f_\alpha(r) \mathfrak v_\lambda^{m\alpha}(t)$,
however, depend on $\phi$ and can be propagated through
Eq.\ \eqref{eomchipmbasis}.
Different azimuthal modes decouple (except, of course, $\pm n$),
because the background does not depend on $\phi$ so that for each
angular basis function, $g_n(\phi) \propto e^{in\phi}$, a set of
coupled evolution equations follows.
%


Since I consider fluctuations above a dynamical background, I must choose
observables affected as little as possible by any mean field motion.
For any changes of the interaction strength, radial breathing
oscillations of the order parameter $\Psi$ are usually excited,
whereas the background stays at rest in azimuthal direction.
I will therefore pick azimuthal density correlations in the following,
where the radial dependence has been integrated out
    \bea
    \hat\varrho(\phi)
    \, = \,
    \int dr r\, \hat \varrho
    \, = \,
    \frac{N}{2\pi}
    + g_m(\phi) \delta\varrho^m_{\lambda} \hat a_\lambda
    + g_m^*(\phi) (\delta\varrho^m_{\lambda})^* \hat a_\lambda^\dagger
    \ea
The measure $r$ has been included for convenience.
The mean azimuthal density, $N/2\pi$, is constant due to
isotropy of potential \eqref{torus},
while the coefficients 
$\delta\varrho^m_{\lambda} = (1/\sqrt\pi) \sum_\alpha
( \mathfrak u_\lambda^{m\alpha} \Re\Psi^\alpha
+ \mathfrak v_\lambda^{m\alpha} \Im\Psi^\alpha )$
of the linear density fluctuations give the nontrivial contributions
of the initial eigenmode $\lambda$ to the density fluctuations with
azimuthal wavenumber $m$, so that the normalized equal-time
density-density correlations follow
    \bea
    \frac{
      \left\langle\hat\varrho(\phi)\hat\varrho(\phi')\right\rangle}
      {\left\langle\hat\varrho(\phi)\right\rangle
        \left\langle\hat\varrho(\phi')\right\rangle}
    \, = \,
    1 +
    \frac{e^{im(\phi - \phi')}}{\sqrt{2\pi}}
    \frac{|\delta\varrho_m|^2}{\varrho_0^2}
    \label{azi_corr}
    \ea
with
$|\delta\varrho_m|^2 = \sum_\lambda|\delta\varrho_\lambda^m|^2/\sqrt{2\pi}$.


\section*{References}
\begin{thebibliography}{499}

\bibitem{Bradley}
A.\ S.\ Bradley,
Phys.\ Rev.\ A {\bf 79}, 033624 (2009). 

\bibitem{Solenov}
D.\ Solenov and D.\ Mozyrsky,
preprint {\sf arXiv:0909.3501}.

\bibitem{Piazza}
F.\ Piazza, L.\ A.\ Collins, and A.\ Smerzi,
Phys.\ Rev.\ A {\bf 80}, 021601 (2009).

\bibitem{Piazza2}
F.\ Piazza, L.\ A.\ Collins, and A.\ Smerzi,
{\sf preprint arXiv:0912.3209}

\bibitem{Jackson}
A.\ D.\ Jackson and G.\ M.\ Kavoulakis,
Phys.\ Rev.\ A {\bf 74}, 065601 (2006).

\bibitem{Tsuchiya}
S.\ Tsuchiya and Y.\ Ohashi,
Phys.\ Rev.\ A {\bf 78}, 013608 (2008).

\bibitem{Tsuchiya2}
S.\ Tsuchiya and Y.\ Ohashi,
Phys.\ Rev.\ A {\bf 79}, 063619 (2009).

\bibitem{Watanabe}
G.\ Watanabe, F.\ Dalfovo, F.\ Piazza, L.\ P.\ Pitaevskii, and
S.\ Stringari,
Phys.\ Rev.\ A {\bf 80}, 053602 (2009).

\bibitem{Didier}
N.\ Didier, A.\ Minguzzi, and F.\ W.\ J.\ Hekking,
Phys.\ Rev.\ A {\bf 79}, 063633 (2009).

\bibitem{Lu-Dac}
M.\ Lu-Dac and V.\ V.\ Kabanov,
Phys.\ Rev.\ B {\bf 79}, 184521 (2009).

\bibitem{Kanamoto}
R.\ Kanamoto, H.\ Saito, and M.\ Ueda,
Phys.\ Rev.\ A {\bf 68}, 043619 (2003).

\bibitem{Kavoulakis}
G.\ M.\ Kavoulakis,
Phys.\ Rev.\ A {\bf 67}, 011601(R) (2003).

\bibitem{Oegren}
M.\ \"Ogren and G.\ M.\ Kavoulakis,
J.\ Low Temp.\ Phys.\ {\bf 154}, 30 (2009).

\bibitem{Salasnich}
L.\ Salasnich, A.\ Parola, and L.\ Reatto,
Phys.\ Rev.\ A {\bf 59}, 2990 (1999).

\bibitem{Javanainen}
J.\ Javanainen, S.\ M.\ Paik, and S.\ M.\ Yoo
Phys.\ Rev.\ A {\bf 58}, 580 (1998).

\bibitem{Benakli}
M.\ Benakli, S.\ Raghavan, A.\ Smerzi, S.\ Fantoni, and
S.\ R.\ Shenoy,
Europhys.\ Lett.\ {\bf 46}, 275 (1999).

\bibitem{Petrosyan}
K.\ G.\ Petrosyan and L.\ You,
Phys.\ Rev.\ A {\bf 59}, 639 (1999).

\bibitem{Andersen}
M.\ F.\ Andersen, C.\ Ryu, P.\ Clad\'e, V.\ Natarajan,
A.\ Vaziri, K.\ Helmerson, and W.\ D.\ Phillips,
Phys.\ Rev.\ Lett.\ {\bf 97}, 170406 (2006).

\bibitem{Ryu}
C.\ Ryu, M.\ F.\ Andersen, P.\ Clad\'e, V.\ Natarajan, K.\ Helmerson,
and W.\ D.\ Phillips,
Phys.\ Rev.\ Lett.\ {\bf 99}, 260401 (2007).

\bibitem{Schnelle}
S.\ K.\ Schnelle, E.\ D.\ van Ooijen, M.\ J.\ Davis,
N.\ R.\ Heckenberg, and H.\ Rubinsztein-Dunlop,
Optics Express {\bf 16}, 1405 (2008).

\bibitem{Wright}
E.\ M.\ Wright, J.\ Arlt, and K.\ Dholakia,
Phys.\ Rev.\ A {\bf 63}, 013608 (2000).

\bibitem{Raman}
C.\ Raman, M.\ K\"ohl, R.\ Onofrio, D.\ S.\ Durfee, C.\ E.\ Kuklewicz,
Z.\ Hadzibabic, and W.\ Ketterle
Phys.\ Rev.\ Lett.\ {\bf 83}, 2502 (1999).

\bibitem{Madison}
K.\ W.\ Madison, F.\ Chevy, W.\ Wohlleben, and J.\ Dalibard,
Phys.\ Rev.\ Lett.\ {\bf 84}, 806 (2000).

\bibitem{Simula}
T.\ P.\ Simula, N.\ Nygaard, S.\ X.\ Hu, L.\ A.\ Collins,
B.\ I.\ Schneider, and K.\ M\o lmer,
Phys.\ Rev.\ A {\bf 77}, 015401 (2008).

\bibitem{Leszczyszyn}
A.\ M.\ Leszczyszyn, G.\ A.\ El, Yu.\ G.\ Gladush, and
A.\ M.\ Kamchatnov,
Phys.\ Rev.\ A {\bf 79}, 063608 (2009).

\bibitem{Unruh}
W.\ G.\ Unruh,
Phys.\ Rev.\ Lett.\ {\bf 46}, 1351 (1981);
Phys.\ Rev.\ D {\bf 51}, 2827 (1995).

\bibitem{Review}
C.\ Barcel\'o, S.\ Liberati, and M.\ Visser,
Living Rev.\ Relativ.\ {\bf 8}, 12 (2005).

\bibitem{Garay}
L.\ J.\ Garay, J.\ R.\ Anglin, J.\ I.\ Cirac, and P.\ Zoller,
Phys.\ Rev.\ Lett.\ {\bf 85}, 4643 (2000).

\bibitem{Jain}
P.\ Jain, A.\ S.\ Bradley, and C.\ W.\ Gardiner,
Phys.\ Rev.\ A {\bf 76}, 023617 (2007).

\bibitem{Leonhardt}
U.\ Leonhardt, T.\ Kiss, and P.\ \"Ohberg,
Phys.\ Rev.\ A {\bf 67}, 033602 (2003).

\bibitem{Recati}
A.\ Recati, N.\ Pavloff and I.\ Carusotto,
preprint {\sf arXiv:0907.4305}.

\bibitem{Carusotto}
I.\ Carusotto, S.\ Fagnocchi, A.\ Recati, R.\ Balbinot, and
A.\ Fabbri,
New J.\ Phys.\ {\bf 10}, 103001 (2008).

\bibitem{Barcelo}
C.\ Barcel\'o, S.\ Liberati, and M.\ Visser,
Phys.\ Rev.\ A {\bf 68}, 053613 (2003).

\bibitem{Giovanazzi}
S.\ Giovanazzi, C.\ Farrell, T.\ Kiss, and U.\ Leonhardt,
Phys.\ Rev.\ A {\bf 70}, 063602 (2004).

\bibitem{Wuester}
S.\ W\"uster and C.\ M.\ Savage,
Phys.\ Rev.\ A {\bf 76}, 013608 (2007).

\bibitem{Wuester2}
S.\ W\"uster,
Phys.\ Rev.\ A {\bf 78}, 021601(R) (2008).

\bibitem{Giovanazzi2}
S.\ Giovanazzi,
Phys.\ Rev.\ Lett.\ {\bf 94}, 061302 (2005).

\bibitem{Balbinot}
R.\ Balbinot, A.\ Fabbri, S.\ Fagnocchi, A.\ Recati, and
I.\ Carusotto,
Phys.\ Rev.\ A {\bf 78}, 021603(R) (2008).

\bibitem{Barcelo3}
C.\ Barcel\'o, A.\ Cano, L.\ J.\ Garay, and G.\ Jannes,
Phys.\ Rev.\ D {\bf 74}, 024008 (2006).

\bibitem{Macher}
J.\ Macher and R.\ Parentani,
Phys.\ Rev.\ A {\bf 80}, 043601 (2009).

\bibitem{Hawking}
S.\ W.\ Hawking,
Nature {\bf 248}, 30 (1974);
Commun.\ Math.\ Phys.\ {\bf 43}, 199 (1975).

\bibitem{Guth}
A.\ Guth,
Phys.\ Rev.\ D {\bf 23}, 347 (1981);
%
A.\ Guth and S.-Y.\ Pi,
Phys.\ Rev.\ Lett.\ {\bf 49}, 1110 (1982).

\bibitem{BirrellDavies}
N.\ D.\ Birrell and P.\ C.\ W.\ Davies,
{\em Quantum fields in curved space},
(Cambridge University Press, Cambridge, UK, 1982).

\bibitem{LiddleLyth}
A.\ R.\ Liddle and D.\ H.\ Lyth,
{\em Cosmological Inflation and Large-Scale Structure},
(Cambridge University Press, Cambridge, UK, 1982).

\bibitem{ich-pra}
M.\ Uhlmann,
Phys.\ Rev.\ A {\bf 79}, 033601 (2009).

\bibitem{ich-njp}
M.\ Uhlmann, Y.\ Xu, and R.\ Sch\"utzhold,
New J.\ Phys.\ {\bf 7}, 248 (2005).

\bibitem{Fedichev}
P.\ O.\ Fedichev and U.\ R.\ Fischer,
Phys.\ Rev.\ Lett. {\bf 91}, 240407 (2003).

\bibitem{Fedichev2}
P.\ O.\ Fedichev and U.\ R.\ Fischer,
Phys.\ Rev.\ A {\bf 69}, 033602 (2004).

\bibitem{Jain2}
P.\ Jain, S.\ Weinfurtner, and M.\ Visser,
Phys.\ Rev.\ A {\bf 76}, 033616 (2007).

\bibitem{Barcelo4}
C.\ Barcel\'o, S.\ Liberati, and M.\ Visser,
Int.\ J.\ Mod.\ Phys.\ D {\bf 12}, 1641 (2003).

\bibitem{Barcelo5}
C.\ Barcel\'o, S.\ Liberati, and M.\ Visser,
Class.\ Quant.\ Grav.\ {\bf 18}, 1137 (2001);
{\em ibid}. {\bf 18}, 3595 (2001).

\bibitem{Kurita}
Y.\ Kurita and T.\ Morinari,
Phys.\ Rev.\ A {\bf 76}, 053603 (2007).

\bibitem{Fischer2}
U.\ R.\ Fischer and R.\ Sch\"utzhold,
Phys.\ Rev.\ A {\bf 70}, 063615 (2004).

\bibitem{Carusotto2}
I.\ Carusotto, R.\ Balbinot, A.\ Fabbri, and A.\ Recati,
Eur.\ Phys.\ J.\ D {\bf 56}, 391 (2010).

\bibitem{Imambekov}
A.\ Imambekov, I.\ E.\ Mazets, D.\ S.\ Petrov, V.\ Gritsev, S.\ Manz,
S.\ Hofferberth, T.\ Schumm, E.\ Demler, and J.\ Schmiedmayer,
Phys.\ Rev.\ A {\bf 80}, 033604 (2009).

\bibitem{Manz}
S.\ Manz, R.\ B\"ucker, T.\ Betz, Ch.\ Koller, S.\ Hofferberth,
I.\ E.\ Mazets, A.\ Imambekov, E.\ Demler, A.\ Perrin,
J.\ Schmiedmayer, and T.\ Schumm,
preprint {\sf arXiv:0911.2376}.


\bibitem{Gangardt}
D.\ M.\ Gangardt and M.\ Pustilnik,
Phys.\ Rev.\ A {\bf 77}, 041604(R) (2008).


\bibitem{Modugno}
M.\ Modugno, C.\ Tozzo, and F.\ Dalfovo,
Phys.\ Rev.\ A {\bf 74}, 061601(R) (2006).

\bibitem{Feshbach}
E.\ Tiesinga, B.\ J.\ Verhaar, and H.\ T.\ C.\ Stoof,
Phys.\ Rev.\ A {\bf 47}, 4114 (1993);
%
S.\ Inouye, M.\ R.\ Andrews, J.\ Stenger, H.-J.\ Miesner,
D.\ M.\ Stamper-Kurn, and W.\ Ketterle,
Nature (London) {\bf 392}, 151 (1998);
%
P.\ Courteille, R.\ S.\ Freeland, D.\ J.\ Heinzen,
F.\ A.\ van Abeelen, and B.\ J.\ Verhaar,
Phys.\ Rev.\ Lett.\ {\bf 81}, 69 (1998).
%
J.\ Weiner, V.\ S.\ Bagnato, S.\ Zilio, and P.\ S.\ Julienne,
Rev.\ Mod.\ Phys.\ {\bf 71}, 1 (1999).
%
S.\ E.\ Pollack, D.\ Dries, M.\ Junker, Y.\ P.\ Chen,
T.\ A.\ Corcovilos, and R.\ G.\ Hulet,
Phys.\ Rev.\ Lett.\ {\bf 102}, 090402 (2009).

\bibitem{PitaevskiiStringari}
L.\ Pitaevskii and S.\ Stringari,
{\em Bose-Einstein Condensation}
(Clarendon, Oxford, 2003).

\bibitem{Bloch}
I.\ Bloch, J.\ Dalibard, and W.\ Zwerger,
Rev.\ Mod.\ Phys.\ {\bf 80}, 885 (2008).

\bibitem{meanfield}
M.\ Girardeau and R.\ Arnowitt,
Phys.\ Rev.\ {\bf 113}, 755 (1959);
%
C.\ Q.\ Gardiner,
Phys.\ Rev.\ A {\bf 56}, 1414 (1997);
%
M.\ D.\ Girardeau,
{\em ibid} {\bf 58}, 775 (1998).

\bibitem{Gross}
E.\ P.\ Gross,
Nuovo Cimento {\bf 20}, 454 (1961);
%
L.\ P.\ Pitaevskii,
Zh.\ Eksp.\ Teor.\ Fiz. {\bf 40}, 646 (1961);
[Sov.\ Phys.\ JETP {\bf 13}, 451 (1961)].

\bibitem{Bogoliubov}
N.\ N.\ Bogoliubov,
J.\ Phys.\ (Moscow) {\bf 11}, 23 (1947);
%
P.\ G.\ de Gennes,
{\em Superconductivity of Metals and Alloys}
(Benjamin, New York, 1966).

\bibitem{nonsymmop}
  Note that the operator $\D$ is not symmetric with respect to the
  usual $L^2 \times L^2$ inner product.
  %
  Although it is possible to introduce a Klein-Gordon type product with
  respect to which $\D$ is self-adjoint \cite{Leonhardt}, this comes
  at the cost of losing positive definiteness (of the inner product).
  %
  I will use the usual $L^2 \times L^2$ scalar product throughout this
  article, i.e., discuss the eigenvalue problem of a non-symmetric
  operator $\D$.
  %
  For real non-symmetric operators, right and left eigenfunctions, i.e.,
  the eigenfunctions of $\D$ and its adjoint $\D^\dagger$, are generally
  distinct but still have the same spectrum with eigenvalue pairs
  $\pm \lambda$ and $\pm \lambda^*$.
  %
  Usually, right (left) eigenfunctions are not mutually orthogonal,
  but an orthogonality relation between left and right eigenfunction
  holds, i.e., any left eigenfunction is orthogonal to all right
  eigenfunctions with different eigenvalues.
  %
  Also, the set of eigenfunctions does not always span the entire
  Hilbert space ($L^2\times L^2$), though this is usually the case.
  %
  It should be noted that in certain situations, e.g., for a homogeneous
  phase $S_0$ of the order parameter $\Psi$ \cite{homphase}, the
  eigenproblem of $\D$ can be mapped onto that of a symmetric operator
  so that the eigenfunctions of $\D$ can be shown to be complete.

\bibitem{homphase}
  For instance, if the background phase is homogeneous, a (possibly
  time-dependent) phase transformation can render $\mathcal C$ zero.
  %
  The operator $\D^2$ will then become diagonal with elements 
$-\mathcal A\mathcal B$ and $-\mathcal B\mathcal A$.
  %
  Since $\mathcal A$ and $\mathcal B$ are both self-adjoint, it is
  possible to define $\sqrt{\mathcal A}$ and $\sqrt{\mathcal B}$ through
  the spectral theorem.
  %
  The eigenvalue problem of $\D^2$ can then be rewritten as
  \bea
    \left(\begin{array}{cc}
      -\sqrt{\mathcal B}\mathcal A\sqrt{\mathcal B} & 0\\
      0 & - \sqrt{\mathcal A}\mathcal B\sqrt{\mathcal A}
    \end{array}\right)
    \left(\begin{array}{c}
      \sqrt{\mathcal B} u_\lambda\\
      \sqrt{\mathcal A} v_\lambda
    \end{array}\right)
    \, = \,
    \lambda
    \left(\begin{array}{c}
      \sqrt{\mathcal B} u_\lambda\\
      \sqrt{\mathcal A} v_\lambda
    \end{array}\right) \,,
    \nonumber
    \ea
  which is that of an symmetric operator.

\bibitem{Hellweg}
D.\ Hellweg, L.\ Cacciapuoti, M.\ Kottke, T.\ Schulte, K.\ Sengstock,
W.\ Ertmer, and J.\ J.\ Arlt,
Phys.\ Rev.\ Lett.\ {\bf 91}, 010406 (2003).

\bibitem{Simsarian}
J.\ E.\ Simsarian, J.\ Denschlag, M.\ Edwards, C.\ W.\ Clark, L.\ Deng,
E.\ W.\ Hagley, K.\ Helmerson, S.\ L.\ Rolston, and W.\ D.\ Phillips,
Phys.\ Rev.\ Lett. {\bf 85}, 2040 (2000).

\bibitem{Mathey}
L.\ Mathey, A.\ Vishwanath, and E.\ Altman,
Phys.\ Rev.\ A {\bf 79}, 013609 (2009).

\bibitem{Negretti}
A.\ Negretti, C.\ Henkel, and K.\ M\o lmer,
Phys.\ Rev.\ A {\bf 78}, 023630 (2008).

\bibitem{Altman}
E.\ Altman, E.\ Demler, and M.\ D.\ Lukin,
Phys.\ Rev.\ A {\bf 70}, 013603 (2004).

\bibitem{Sykes}
A.\ G.\ Sykes, D.\ M.\ Gangardt, M.\ J.\ Davis, K.\ Viering,
M.\ G.\ Raizen, and K.\ V.\ Kheruntsyan,
Phys.\ Rev.\ Lett.\ {\bf 100}, 160406 (2008).

\bibitem{Esteve}
J.\ Est\`eve, J.-B.\ Trebbia, T.\ Schumm, A.\ Aspect, C.\ I.\ Westbrook,
and I.\ Bouchoule,
Phys.\ Rev.\ Lett.\ {\bf 96}, 130403 (2006).

\bibitem{Schellekens}
M.\ Schellekens, R.\ Hoppeler, A.\ Perrin, J.\ Viana Gomes,
D.\ Boiron, A.\ Aspect, C.\ I.\ Westbrook,
Science {\bf 310}, 648 (2005).

\bibitem{Ralf}
R.\ Sch\"utzhold,
Phys.\ Rev.\ Lett.\ {\bf 97}, 190405 (2006).

\bibitem{Cherny}
A.\ Yu.\ Cherny and J.\ Brand,
Phys.\ Rev.\ A {\bf 79}, 043607 (2009).

\bibitem{Petrov}
D.\ S.\ Petrov and G.\ V.\ Shlyapnikov,
Phys.\ Rev.\ A {\bf 64}, 012706 (2001).


\bibitem{Bader}
P.\ Bader and U.\ R.\ Fischer,
Phys.\ Rev.\ Lett.\ {\bf 103}, 060402 (2009).

\bibitem{Fischer}
U.\ R.\ Fischer and P.\ Bader,
preprint {\sf arXiv:0908.2026}.




\end{thebibliography}
\end{document}